\documentclass[english,english, review]{elsarticle}
\usepackage[T1]{fontenc}
\usepackage[latin9]{inputenc}
\usepackage{color}
\usepackage{array}
\usepackage{rotfloat}
\usepackage{amsmath}
\usepackage{amsthm}
\usepackage{amssymb}
\usepackage{graphicx}

\makeatletter

\providecommand{\tabularnewline}{\\}

  \theoremstyle{definition}
  \newtheorem*{example*}{\protect\examplename}

\newdefinition{rmk}{Remark}
\newproof{pf}{Proof}
\newproof{pot}{Proof of Theorem \ref{thm2}}

\@ifundefined{showcaptionsetup}{}{%
 \PassOptionsToPackage{caption=false}{subfig}}
\usepackage{subfig}
\makeatother

\usepackage{babel}
  \providecommand{\examplename}{Example}

\begin{document}

\title{Evidential positive opinion influence measures for Viral Marketing }

\author[ISG]{Siwar Jendoubi\corref{mycorrespondingauthor}}
\cortext[mycorrespondingauthor]{Corresponding author}
\ead{jendoubi.siwar@yahoo.fr}
\author[IRISA]{Arnaud Martin}

\address[ISG]{LARODEC, ISG Tunis, University of Tunis, Avenue de la Libert\'e, Cit\'e Bouchoucha, Le Bardo 2000, Tunisia}

\address[IRISA]{DRUID, univ Rennes, CNRS, IRISA, Rue E. Branly, 22300 Lannion, France}

\begin{abstract}
The Viral Marketing is a relatively new form of marketing that exploits
social networks to promote a brand, a product, \textit{etc}. The idea
behind it is to find a set of influencers on the network that can
trigger a large cascade of propagation and adoptions. In this paper,
we will introduce an evidential opinion-based influence maximization
model for viral marketing. Besides, our approach tackles three opinions
based scenarios for viral marketing in the real world. The first scenario
concerns influencers who have a positive opinion about the product.
The second scenario deals with influencers who have a positive opinion
about the product and produce effects on users who also have a positive
opinion. The third scenario involves influence users who have a positive
opinion about the product and produce effects on the negative opinion
of other users concerning the product in question. Next, we proposed
six influence measures, two for each scenario. We also use an influence
maximization model that the set of detected influencers for each scenario.
Finally, we show the performance of the proposed model with each influence
measure through some experiments conducted on a generated dataset
and a real world dataset collected from Twitter.
\end{abstract}
\maketitle
\section{Introduction}

The unprecedented growth of social networks made the marketers revalue
their strategies in viral marketing. Viral marketing exploits existing
social networks and sends viral marketing messages related to a company,
brand or product by influencing friends who will recommend intentionally
or unintentionally the product to other friends and many individuals
will probably adopt it. Hotmail rapidly adopted a viral marketing
strategy and reported a significant rise of their business from influence
propagation in social networks. Hotmail gained 18 million users in
12 months, spending only \$50,000 on traditional marketing \citep{Jurvetson2000},
while Gmail rapidly gained users although referrals are the only way
to sign up.

The phenomenon of influence propagation through social networks has
been attracting a great body of research works \citep{Kempe03,Goyal12,Barbieri2013,Narayanam2014}.
However, existing influence maximization approaches assume that all
network users and influencers have a positive opinion and the works
\citep{Zhang2013,Chen2010} that consider the user's positive opinion
assume the availability of positive influence probabilities. Whereas,
a key function of social networks is sharing, by enabling users to
express their opinions about a product or trend of news by means of
posts, shared posts, likes/dislikes, or comments on friends' posts.
Such opinions are propagated to other users and might exert an either
positive or negative influence on them. For example, if some friends
have shown any positive (or negative) comments against a product or
news, one will have a similar feeling regardless of their own opinion.
Then, the influencer opinion is beneficial for the product if it is
positive. However, such an opinion is harmful if it is negative. In
fact, according to the social psychology literature \citep{Taylor1991,Baumeister2001}
the information negativity (negative events, ideas, news, \textit{etc.})
is always stronger than positivity (positive events, ideas, news,
\textit{etc.}). Consequently, the consideration of the user's opinion
in the influence maximization process is crucial for a given marketing
campaign. 

The influence maximization (IM) in an online social network (OSN)
presents two main challenges: the first challenge is about data imprecision
and uncertainty. In fact, social interactions are not always precise
and certain. Besides, the OSN API (Application Programming Interface)
allows only a limited access to their data which generates more imprecision
and uncertainty for the social network analysis fields. Then, if we
ignore this imperfection of the data, we may be confronted with erroneous
analysis results. The second challenge is about the diversity of influence
markers and parameters, \textit{e.g.} the opinion, the user's activity,
the position in the network, \textit{etc}. Indeed, it is important
to combine all of them to obtain a global influence measure that considers
all these parameters, the data imperfection and the conflict that
may exist between influence markers. Consequently, it is necessary
to resolve these challenges in order to obtain better influence maximization
results. For these purposes, we propose the theory of belief functions
\citep{Dempster67a,Shafer76} as a solution. In fact, it was widely
applied in many similar situations and it was efficient. Furthermore,
this theory was used for analyzing social networks \citep{Wei13,Gao13,Jendoubi14a,Jendoubi2015,Jendoubi2016a,Jendoubi2017,Jendoubi2018}.

In this paper, we consider the user's opinion about the product and
we contribute with three opinion-based scenarios of viral marketing.
To the best of our knowledge, this is the first work that refines
the influencers detection process to be up to the expectations of
marketers. In fact, we introduce technical solutions that fit more
with purposes of their campaigns. In the first scenario, we investigate
influencers who have a positive opinion about the product. The second
scenario looks for influencers who have a positive opinion about the
product and exert more influence on users with a positive opinion.
Finally, the third scenario searches to detect influence users who
have a positive opinion about the product and exert more influence
on influence users having a negative opinion about the product. For
each scenario, we define two evidential influence measures. We use
the theory of belief functions to estimate the influence considering
data imperfection. 

To the best of our knowledge, the work conducted in this paper is
the first to achieve the following contributions: 1) we consider the
real user's opinion that is estimated from the posted messages, 2)
we introduce three opinion-based scenarios of viral marketing which
gives flexible solutions that go with marketers expectations, 3) we
propose two evidential opinion-based influence measures for each scenario,
4) we use the theory of belief functions to remedy the data imperfection
problems, to fuse influence markers and user opinion efficiently,
5) we maximize the influence in an OSN using the proposed measures,
6) we conduct many experiments on generated data and real-world data
collected from Twitter to show the performance of the proposed measures
with the used influence maximization model.

This paper is organized as follows: Section 2 reviews some relevant
existing works, Section 3 deals with backgrounds of the theory of
belief functions that are used in this paper. A few lines later, Section
4 focuses our attention on the introduction of the proposed three
opinion-based scenarios of viral marketing. A set of experiments is
presented and discussed in Section 5.

\section{Related work\label{sec:Related-work}}

The influence maximization (IM) is a very common problem that has
been widely studied since its introduction \citep{Domingos01,Kempe03,LiYung2011}.
It searches to select a set of $k$ users in the social network that
could adopt the product and trigger a large cascade of adoptions and
propagation through the ``word of mouth'' effect. The selected users
are commonly called influencers. In the literature, we find many solutions
for the IM problem. In this section, we present an overview of the
state of the art.

\subsection{Influence maximization: an overview\label{subsec:Influence-maximization}}

The influence maximization (IM) is the problem of finding a set $S$
of $k$ influence users that can trigger a large cascade of propagation
and adoption. The set $S$ is called seed set. A very common application
of this problem is the viral marketing \citep{LiYung2011}. Indeed,
in a viral marketing campaign, the marketer wishes to propagate propaganda
and to make it go viral. Then, he needs a set of influencers to be
the triggers of the campaign. Those influencers will start the propagation
process. Kempe \textit{et al.} \citep{Kempe03} were the first to
define the problem of detecting influencers as a maximization problem.
Besides, they proved the NP-Hardness of this problem. Given a social
network $G=\left(V,E\right)$, where $V$ is the set of nodes and
$E$ is the set of links, they defined $\sigma_{M}\left(S\right)$
as the expected number of influenced nodes in the network. Next, Kempe\textit{
et al.} \citep{Kempe03} showed that $\sigma_{M}\left(S\right)$ is
monotone, \textit{i.e.} $\sigma_{M}\left(S\right)\leq\sigma_{M}\left(T\right)$
whenever $S\subseteq T\subseteq V$, and sub-modular, \textit{i.e.}
$\sigma_{M}\left(S\cup\left\{ x\right\} \right)-\sigma_{M}\left(S\right)\geq\sigma_{M}\left(T\cup\left\{ x\right\} \right)-\sigma_{M}\left(T\right)$
whenever $S\subseteq T\subseteq V$ and $x\in V$. In the influence
maximization step, \textit{i.e.} maximizing $\sigma_{M}\left(S\right)$,
the authors proposed the greedy algorithm with the Monte Carlo simulation.

To calculate $\sigma_{M}\left(S\right)$, \citep{Kempe03} used two
existing propagation models: the\textit{ Linear Threshold Model (LTM)
}\citep{Granovetter78} and the \textit{Independent Cascade Model
(ICM)} \citep{Goldenberg01}. A node $v$ is \textit{active} if it
receives the propagated information and accepts it, elsewhere, it
is \textit{inactive}. An inactive vertex can change its status to
active. The ICM model considers that each active vertex $u$ has only
one chance to activate its neighbors. Indeed, when a given node $u$
becomes active at $t$, it will try to activate its inactive neighbor
$v$ at $t+1$ with a success probability $p\left(u,v\right)$ (parameter
of the system). The \textit{Weighted Cascade (WC)} is a special case
of ICM where 
\begin{equation}
p\left(u,v\right)=\frac{1}{D_{u}}
\end{equation}
such that $D_{u}$ is the degree of the node $u$. The LTM model has
another activation condition based on a node \textit{threshold} $\theta_{u}$.
It associates a \textit{weight} $\omega\left(u,v\right)$ to each
link $\left(u,v\right)$ and a threshold $\theta_{u}$ to each node
$u$. A given inactive node $u$ changes its status to active if the
following condition is true\textit{:} 
\begin{equation}
\sum_{v}\omega\left(u,v\right)\geq\theta_{u}
\end{equation}
where $\theta_{u}\in\left[0,1\right]$ is a random uniform variable
and it is defined as the tendency of a given node $u$ to adopt the
information when its neighbors do.

The work of \citep{Xiang2013,Liu2017} studied the influence propagation
using the Page Rank algorithm. In fact, they studied the connection
between the social influence and Page Rank and they found that Page
Rank information can be helpful to find influencers. Xiang \textit{et
al.} \citep{Xiang2013} introduced a linear social influence model
that approximates ICM. Next, they showed that the Page Rank algorithm
is a special case of their algorithm. The work of Liu \textit{et al.}
\citep{Liu2017} is an extension of \citep{Xiang2013} in which they
consider nodes' prior knowledge to better identify influencers. We
find also, the work of \citep{Yang2018} that introduced a marketing
campaign recommender system udapted for Business to Business (B2B).
Their work is interesting as it considers ``the temporal behavior
patterns in the B2B buying processes'' \citep{Yang2018}. We notice
that these solutions do not consider the user opinion which implicates
a hidden assumption that considers all influencers as positive.

Another interesting influence maximization model was introduced by
Goyal \textit{et al.} \citep{Goyal12}. It is a data-based model called
credit distribution (CD). It takes two main inputs which are the network
structure and propagation log. It uses these inputs to estimate the
influence. In fact, the influence spread function is defined as the
total influence credit given to a set of users $S$ from the whole
network. The idea behind this algorithm is when an action $a$ propagates
from a user $v$ to a user $u$, a direct influence credit $\gamma_{v,u}\left(a\right)$
is given to $v$. Furthermore, a credit amount is given to the predecessors
of $v$. The credit distribution algorithm starts by scanning the
action log $L$ (the set of tuples $\left(User,\,Action,\,Time\right)$
where $\left(v,a,t\right)\in L$ when $v$ performed the action $a$
at time $t$) and estimates the total influence credit given to $v$
for influencing $u$ on the action $a$, $\Gamma_{v,u}\left(a\right)$.
Initially, we have $S=\emptyset$. Next, CD runs up the CELF algorithm
to detect the vertex having the maximum marginal gain and so on until
having $k$ nodes in $S$. More details can be found in \citep{Goyal12}.

As the work of \citep{Goyal12}, we consider past propagation for
influence maximization. However, the proposed approach differs from
their solution. In fact, we consider the user's opinion about the
product and the opinions of the influencer neighbors. 

\subsection{Opinion-based influence maximization\label{subsec:Opinion-based-influence-maximiza}}

In this section, we are mainly interested in works that incorporate
the user's opinion in the influence maximization process which is
a relatively new idea. In fact, the user's opinion is a critical factor
in marketing and social science. In the social psychology literature,
the concept of positive-negative opinion asymmetry was largely studied
\citep{Taylor1991,Baumeister2001}. These works agreed on the fact
that negativity (negative events, ideas, news, \textit{etc.}) is always
stronger than positivity (positive events, ideas, news, \textit{etc.}).
This fact was, also, shown in marketing science like the work of Cheung
and lee \citep{Cheung2008} that studied the impact of the negative
electronic word of mouth (eWoM) on online shops and they found that
``negative eWoM has a significantly larger impact on consumer trust
and intention to the online shop''. These works prove the importance
of the opinion in the influence maximization process and especially
the importance of selected seed's opinion.

Chen \textit{et al.} \citep{Chen2010} extended the influence propagation
model of Kempe \textit{et al.} \citep{Kempe03} and incorporated the
negative opinions and their propagation into the influence maximization
process. On an interesting work, Zhang\textit{ et al.} \citep{Zhang2013}
proposed the \textit{Opinion-based Cascading (OC)} model that takes
positive opinions of users into consideration. They used the OC model
to maximize the positive influence by considering the user's opinion
and the change of the opinion. They showed that the objective function
of the OC model is no longer submodular. Besides, they proved the
NP-Hardness of their model. Then, they proposed an approximation of
the maximization results in a polynomial time. In a first step, OC
ignores all users that have a small potential marginal gain that is
defined as: 
\begin{eqnarray}
PMG\left(v\right) & = & Op\left(v\right)+\sum_{u\in N_{out}^{a}\left(v\right)}\left(Op\left(u\right)+Op\left(v\right).w\left(v,u\right)\right)\nonumber \\
 & + & \sum_{u\in N_{out}^{ia}\left(v\right)}\frac{w\left(v,u\right)}{\theta_{u}}\left(Op\left(u\right)+Op\left(v\right).w\left(v,u\right)\right)
\end{eqnarray}
where $Op\left(v\right)$ defines the opinion indicator of $v$ such
that $Op\left(v\right)=0$ means that $v$ has a neutral opinion,
$Op\left(v\right)>0$ indicates that the opinion is positive, $Op\left(v\right)<0$
the opinion of $v$ is negative. The sets $N_{out}^{a}\left(v\right)$
and $N_{out}^{ia}\left(v\right)$ are respectively, the sets of $v$'s
active and inactive out-neighbors,\textit{ i.e.} the set of destinations
of directed links having $v$ as source. The parameter $\theta_{u}$
defines the activation probability of $u$. Finally, $w\left(v,u\right)$
is the weight associated to the edge $\left(v,u\right)$. In the next
step, OC iterates until getting $k$ seed nodes. In each iteration,
the algorithm updates the activation status according to the condition
$\sum_{u\in N_{in}^{a}\left(v\right)}w\left(u,v\right)\geq\theta_{v}$
where $N_{in}^{a}\left(v\right)$ is the set of active in-neighbors
of $v$. Also, it updates the opinion value of each user according
to his previously activated neighbors as:

\begin{equation}
Op\left(v\right)=Op\left(v\right)+\sum_{u\in N_{in}^{a}\left(v\right)}\left(Op\left(u\right).w\left(u,v\right)\right)\label{eq:opinionOc}
\end{equation}
Then, OC chooses the user that still in the top of the potential list.
Li \textit{et al.} \citep{Li2014} considered not only the friendship
relations but also foe relations in the influence maximization problem.
They used a signed network, \textit{i.e.} positive relations $\left(+\right)$
to model friendship or trust and negative relations $\left(-\right)$
to model foes or distrust. 

These recent works assumed that positive and negative influence probabilities
are known and given to the influence maximization algorithm as input.
This is obviously not the case in real-world propagation. Indeed,
some preprocessing is needed to close the gap between the model and
the real data \citep{Goyal12}. Furthermore, we tested the opinion-based
cascading model on real world data and with estimated opinion values.
According to these experiments, the opinion-based cascading model
detected several inactive and isolated users, even, the mean positive
opinion of the selected influencers is less than 50\% which is not
a satisfying result. The reader can refer to Section \ref{sec:Experiments}
for more details. These tests confirmed the need for efficient solutions
that deals with these drawbacks.

\subsection{Influence and evidence theory}

In the literature, there are some recent works that use the theory
of belief functions to model the uncertainty while measuring user
influence in online social networks. In this section, we will present
a brief review of the works that we find close to our work.

Authors in \citep{Wei13} presented an evidential centrality (EVC)
measure. EVC is the result of the combination of two BBAs distributions
on the frame $\left\{ high,\,low\right\} $. The first BBA is used
to measure the degree centrality and the second one is used to measure
the strength centrality of the node. The work of \citep{Gao13} proposed
a centrality measure with a same spirit as EVC. In fact, they modified
the EVC measure according to the actual degree of the node instead
of following the uniform distribution, also, they extended the semi-local
centrality measure \citep{Chen12} to be used with weighted networks.
Their centrality measure is the result of the combination of the modified
EVC and the modified semi-local centrality measure. The work of \citep{Gao13}
is similar to the work of \citep{Wei13} in that, they used the same
frame of discernment, their approaches are structure based and they
choose the influential nodes to be top-1 ranked nodes according to
the proposed centrality measure.

Our work is like the work of \citep{Wei13} and \citep{Gao13} in
that we used the same BBA estimation mechanism of the influence BBA.
More details about the influence estimation process we use can be
found in \citep{Jendoubi2017}. Nevertheless, our work is different
from their work in that we consider the user's opinion and we maximized
the influence. In fact, the importance of the opinion parameter in
the viral marketing campaign encouraged us to consider it in the influence
maximization process. 

\section{Background\label{sec:Background}}

In this section, we will introduce some concepts of the theory of
belief functions that are used in this paper. This theory is also
called Dempster-Shafer theory or evidence theory. Dempster \citep{Dempster67a}
was the first to introduce it through what he called \textit{Upper
and Lower probabilities}. Later, \textit{Shafer} published his book
\textquotedbl{}A mathematical theory of evidence\textquotedbl{} \citep{Shafer76}
in which he defined the basics of the evidence theory. This theory
is recommended to process the imprecise and uncertain data. Furthermore,
it allows to achieve more precise, reliable and coherent information.

Let $\Omega=\left\{ d_{1},d_{2},...,d_{n}\right\} $ be a \textit{frame
of discernment}. The basic belief assignment (BBA), $m^{\Omega}$,
is a function that defines the belief on $\Omega$, it is defined
as: 
\begin{eqnarray}
2^{\Omega} & \rightarrow & \left[0,1\right]\nonumber \\
A & \mapsto & m^{\Omega}\left(A\right)
\end{eqnarray}
where $2^{\Omega}=\left\{ \emptyset,\left\{ d_{1}\right\} ,\left\{ d_{2}\right\} ,\left\{ d_{1},d_{2}\right\} ,...,\left\{ d_{1},d_{2},...,d_{n}\right\} \right\} $
is the set of all subsets of $\Omega$ called \textit{power set}.
The mass $m^{\Omega}\left(A\right)$ is the value assigned to the
subset $A\subseteq\Omega$ and it must respect: 
\begin{equation}
\sum_{A\subseteq\Omega}m^{\Omega}\left(A\right)=1\label{eq:mass}
\end{equation}
In the case where we have $m^{\Omega}(A)>0$, $A$ is called focal
element of $m^{\Omega}$. \textit{A simple mass function or simple
BBA} has at most two focal elements among them $\Omega$. Let $\alpha\in\left[0,1\right]$,
the simple BBA $m^{\Omega}$ is defined as: 
\begin{equation}
\begin{cases}
m^{\Omega}\left(A\right) & =1-\alpha,\,A\subseteq\Omega\\
m^{\Omega}\left(\Omega\right) & =\alpha
\end{cases}\label{eq:Simple BBA}
\end{equation}
\begin{example*}
Let's consider the well-known example of the murder of Mr. Jones introduced
by \citep{Smets94a}. Big Boss has a team of assassins composed of
three members who are Peter, Paul and Mary. We need to define a frame
of discernment that contains all possible assassins: $\Omega$ is
formed by: Peter (Pe), Paul (Pa) and Mary (Ma), $\Omega=\left\{ Pe,\,Pa,\,Ma\right\} $,
and its corresponding power set is: 
\begin{equation}
2^{\Omega}=\left\{ \emptyset,\left\{ Pe\right\} ,\left\{ Pa\right\} ,\left\{ Pe,Pa\right\} ,\left\{ Ma\right\} ,\left\{ Pe,Ma\right\} ,\left\{ Pa,Ma\right\} ,\left\{ Pe,Pa,Ma\right\} \right\} 
\end{equation}
Big Boss selected a killer from his team using a dice, if he obtains
an even number, then the killer is a female, else, the killer is a
male. Let's help the Judge to find the murder. We know that Mr. Jones
was murdered and the sex of the murder was selected through a dice.
However, there is no information about the choice between Peter and
Paul in the case of an odd number.

Knowing this information, we can define the following BBA on $\Omega$:\\
 $m_{1}^{\Omega}\left(\left\{ Pe,Pa\right\} \right)=0.5$ and $m_{1}^{\Omega}\left(\left\{ Ma\right\} \right)=0.5$.
\hfill$\square$
\end{example*}
The theory of belief functions presents many combination rules that
are used to fuse pieces of information. The first combination rule
was introduced by Dempster \citep{Dempster67a} and it is called\textbf{
}\textit{Dempster's rule} of combination. It fuses two distinct mass
distributions into a normalized one, \textit{i.e.} a BBA $m^{\Omega}$
is said to be normal if $m^{\Omega}\left(\emptyset\right)=0$. Let
$m_{1}^{\Omega}$ and $m_{2}^{\Omega}$ be two mass distributions
defined on $\Omega$, Dempster's rule is defined as: 
\begin{equation}
m_{1\oplus2}^{\Omega}\left(A\right)=\begin{cases}
\frac{\sum_{B\cap C=A}m_{1}^{\Omega}\left(B\right)m_{2}^{\Omega}\left(C\right)}{1-\sum_{B\cap C=\emptyset}m_{1}^{\Omega}\left(B\right)m_{2}^{\Omega}\left(C\right)}, & A\subseteq\Omega\setminus\left\{ \emptyset\right\} \\
0 & if\,A=\emptyset
\end{cases}\label{eq:dempster rule}
\end{equation}
\begin{example*}
Table (\ref{tab:Combination-rules-example}) is an example of the
Dempster's rule of combination. \hfill$\square$

\begin{table}
\caption{Dempster's rule example\label{tab:Combination-rules-example}}
\centering{}%
\begin{tabular}{|c|c|c|c|}
\cline{2-4} 
\multicolumn{1}{c|}{} & \textbf{$m_{1}^{\Omega}$} & \textbf{$m_{2}^{\Omega}$} & Dempster's rule\tabularnewline
\hline 
\textbf{$\emptyset$} & 0 & 0 & 0\tabularnewline
\hline 
\textbf{$\left\{ Pe\right\} $} & 0 & 0 & 0\tabularnewline
\hline 
\textbf{$\left\{ Pa\right\} $} & 0 & 0 & 0.1765\tabularnewline
\hline 
\textbf{$\left\{ Pe,Pa\right\} $} & 0.5 & 0.3 & 0.4118\tabularnewline
\hline 
\textbf{$\left\{ Ma\right\} $} & 0.5 & 0 & 0.4118\tabularnewline
\hline 
\textbf{$\left\{ Ma,Pe\right\} $} & 0 & 0 & 0\tabularnewline
\hline 
\textbf{$\left\{ Ma,Pa\right\} $} & 0 & 0.3 & 0\tabularnewline
\hline 
\textbf{$\Omega$} & 0 & 0.4 & 0\tabularnewline
\hline 
\end{tabular}
\end{table}
\end{example*}

\section{Proposed opinion-based influence measures}

The user's opinion plays a main role in the viral marketing. In fact,
if a user $A$ shares his negative opinion about a product $x$, then
all users that will receive $A$ opinion will have, at least, some
doubt about $x$ and that in the case in which $A$ is not an influencer
for them. In the case where $A$ is an influence user, then his negative
opinion will be harmful for the product $x$. This fact encouraged
us to propose new influence measures for online social networks that
consider the user's opinion. Besides, we introduce three opinion-based
scenarios of viral marketing that will be helpful for the marketer.
In fact, this work offers more flexibility to the marketer to build
his viral marketing campaign according to his purposes as we explain:
\begin{enumerate}
\item First scenario, \textit{Positive influencers}: in this scenario we
look for influencers who have a positive opinion about the product.
It is useful for marketers who are looking for positive influencer
spreaders. In this case, the marketer may want to avoid influencers
that have a negative opinion, and to target only influence users having
a positive opinion. A solution for this scenario was published in
\textit{Jendoubi et al.} \citep{Jendoubi2016a}.
\item Second scenario, \textit{Positive influencers influencing positive
users}: the purpose in this scenario is to find positive influencers
that exert more influence on users having a positive opinion about
the product. It is destined to marketers who are interested in influence
users that have a positive opinion about the product and that exert
more influence on users having a positive opinion too. In such a case,
the marketer may want to boost the probability of success of his viral
marketing campaign.
\item Third scenario, \textit{Positive influencers influencing negative
users}: the main goal of this scenario is to detect positive influencers
that exert more influence on users having a negative opinion about
the product. It is useful for marketers who are looking for influence
users that have a positive opinion about the product and that exert
more influence on users having a negative opinion. The marketing strategy
here may be, for example, to try to gain more customers by changing
the opinion of users that have a negative opinion.
\end{enumerate}
The set of positive influencers may include positive influencers influencing
positive users and positive influencers influencing negative users.
However, the second and the third scenarios have a more specific selection
criteria as they have a condition on the neighbor\textquoteright s
opinions of the positive influencer. 

In this section, we introduce the opinion polarity estimation process.
Next, we present two influence measures for each scenario. Finally,
we introduce the influence spread function. 

\subsection{Opinion polarity estimation\label{subsec:User's-opinion-estimation}}

In this paper, we consider the user's opinion about the product in
the maximization process. For this purpose, we need to estimate this
opinion. First, we start by estimating the opinion polarity of each
message, then, we take the user's opinion as the mean opinion of his
posted messages. We remark that the message may depend on the considered
online social network (OSN). For example, it can be a tweet, a retweet
or a reply if the OSN is Twitter\footnote{Twitter allows access to the public content through Twitter API},
it can be a wall post or a comment if the OSN is Facebook\footnote{Facebook allows access to pages and groups content. Besides, it is
possible to collect the user's data, but the user's permission is
needed. }, it is a post activity or a comment in GooglePlus\footnote{GooglePlus allows access to its data through an API.}.
Next, we explain the process we used to estimate the opinion expressed
in a given message.

To estimate the opinion polarity of each given message, we used existing
tools that are dedicated for this purpose. We followed the following
process:
\begin{enumerate}
\item We deleted URLs and special characters to simplify the estimation
task. 
\item The second step is called part of speech tagging its goal is to attribute
a label (noun, adjective, verb, adverb) to each word in the message.
For such a purpose, we used the java library ``Stanford part-of-speech
Tagger''\footnote{http://nlp.stanford.edu/software/tagger.shtml}.
We note that the Stanford tagger allows to train a part-of-speech
tagger model using an annotated data corpus. However, it is also possible
to use an available part-of-speech tagger model like the ``GATE Twitter
part-of-speech tagger''\footnote{https://gate.ac.uk/wiki/twitter-postagger.html}
that was designed for Twitter and that is able to achieve about 91\%
of accuracy.
\item We use the SentiWordNet 3.0 \citep{Baccianella2010} dictionary to
get the polarity of each word (positive, negative and objective polarity)
in the message according to its tag (result of the step 2). The result
of this step is a probability distribution defined on $\Theta=\left\{ Pos,\,Neg,\,Neut\right\} $
for each word in the message.
\item To compute the global polarity of the tweet we take the mean probability
distribution of words probability distributions that compose the message.
\end{enumerate}
\begin{example*}
Let us consider the message that says: ``\textit{Smartphones are
good but complicated}''. We want to estimate its opinion polarity
using the explained process. The given message did not contain a URL
or a special character, then the first step is not needed. Table \ref{tab:Computing-the-opinion}
presents the second and the third steps. Finally, the polarity of
the given message is:

\begin{eqnarray}
\textrm{Pr}^{\Theta}\left(Pos\right) & = & 0.152\\
\textrm{Pr}^{\Theta}\left(Neg\right) & = & 0.125\\
\textrm{Pr}^{\Theta}\left(Neut\right) & = & 0.723
\end{eqnarray}

\begin{table}
\caption{Computing the opinion polarity of a message\label{tab:Computing-the-opinion}}
\centering{}%
\begin{tabular}{|c|c|c|c|c|c|}
\cline{2-6} 
\multicolumn{1}{c|}{} & Smartphones & are & good & but & complicated\tabularnewline
\hline 
Part of speech & noun & verb & adj & conjunction & adj\tabularnewline
\hline 
Positive opinion & 0 & 0 & 0.635 & 0 & 0.125\tabularnewline
\hline 
Negative opinion & 0 & 0 & 0.001 & 0 & 0.625\tabularnewline
\hline 
Neutral opinion & 1 & 1 & 0.364 & 1 & 0.25\tabularnewline
\hline 
\end{tabular}
\end{table}
\hfill$\square$
\end{example*}

\subsection{User opinion-based influence estimation\label{subsec:User's-opinion-based}}

Given a social network $G=\left(V,E\right)$, where $V$ is the set
of nodes and $E$ is the set of links, a frame of discernment expressing
opinion $\Theta=\left\{ Pos,\,Neg,\,Neut\right\} $, $Pos$ for positive,
$Neg$ for negative and $Neut$ for neutral and a probability distribution
$\Pr_{u}^{\Theta}$ defined on $\Theta$ that expresses the opinion
of the user $u\in V$ about the product. We transform the opinion
probability distribution $\Pr_{u}^{\Theta}$ to a mass distribution
$m_{u}^{\Theta}$ to consider the uncertainty that may exist in the
user's opinion. We create two simple mass distributions for $\Pr_{u}^{\Theta}\left(Pos\right)$
and $\Pr_{u}^{\Theta}\left(Neg\right)$. In fact, we take $\alpha$,
in formula \ref{eq:Simple BBA} of Section \ref{sec:Background},
equals to $\Pr_{u}^{\Theta}\left(Pos\right)$ for the first BBA and
$\Pr_{u}^{\Theta}\left(Neg\right)$ for the second one. After this
step, we obtain two BBAs expressing the user's positive and negative
opinion respectively. 

Let us define a frame of discernment expressing influence and passivity
$\Omega=\left\{ I,P\right\} $, $I$ for influencers and $P$ for
passive users, and a basic belief assignment (BBA) function \citep{Shafer76},
$m_{\left(u,v\right)}^{\Omega}$, defined on $\Omega$ that expresses
the influence that exerts the user $u$ on the user $v$. An estimation
process for $m_{\left(u,v\right)}^{\Omega}$ was introduced in \citep{Jendoubi2017}.
The influence measure introduced in \citep{Jendoubi2017} does not
consider the user's opinion. In fact, it has an implicit assumption
that all network users have a positive opinion about the product which
is an unrealistic assumption. In the following, an estimation example
of $m_{\left(u,v\right)}^{\Omega}$, this example summarizes the process
described in \citep{Jendoubi2017}.
\begin{example*}
Let us consider two users $u$ and $v$ in the network. To estimate
the influence that exerts $u$ on $v$, we need to define a set of
measurable influence indicators and/or behaviors. For example, the
strength of the relationship between $u$ and $v$, the number of
times $v$ shares $u$'s messages, etc. In the next step, we estimate
a BBA on $\Omega$ for each defined indicator. For this purpose, the
estimation process defined in the work of Wei \textit{et al.} \citep{Wei13}
can be used. Let us consider that we have two influence indicators
and let $m_{1}^{\Omega}$ and $m_{2}^{\Omega}$ be their respective
BBAs: $m_{1}^{\Omega}\left(I\right)=0.5$, $m_{1}^{\Omega}\left(P\right)=0.3$,
$m_{1}^{\Omega}\left(\left\{ I,P\right\} \right)=0.2$, $m_{2}^{\Omega}\left(I\right)=0.4$,
$m_{2}^{\Omega}\left(P\right)=0.5$ and $m_{2}^{\Omega}\left(\left\{ I,P\right\} \right)=0.1$.
To estimate $m_{\left(u,v\right)}^{\Omega}$ we combine the BBAs of
all indicators using the Dempster's rule (equation \ref{eq:dempster rule}
of Section \ref{sec:Background}). Then $m_{\left(u,v\right)}^{\Omega}=m_{1}^{\Omega}\oplus m_{2}^{\Omega}$.\hfill$\square$
\end{example*}
In the next step, we define two influence measures for each opinion
based scenario. Besides, we use the Dempster's rule of combination
for BBAs fusion (equation \ref{eq:dempster rule} of Section \ref{sec:Background})
to obtain $m_{u}^{\Theta}$ that expresses the opinion of $u$.

\subsubsection{Positive opinion influencer \label{subsec:First-scenario}}

The goal in the first scenario is to detect social influencers who
have a positive opinion about the product. In fact, we search to avoid
negative influencers, because targeting these users may have a harmful
effect on the Viral Marketing campaign. For example, the marketer
wants to promote his product in an online social network. First, he
starts by identifying a set of influencers in the network that maximizes
the total influence. Second, he contacts them and tries to convince
them to do some advertising for his product. He may give the influencers
a free product or a discounting in order to encourage them more to
do the advertising. If by chance he falls on some influencers that
do not like his product, what would be their reaction in such a case?
Then we propose to avoid negative influencers by detecting and targeting
positive influencers. As defined above, the mass value $m_{\left(u,v\right)}^{\Omega}\left(I\right)$
measures the influence of $u$ on $v$ but without considering the
opinion of $u$ about the product. We define the positive opinion
influence of $u$ on $v$ as the positive proportion of $m_{\left(u,v\right)}^{\Omega}\left(I\right)$
and we propose two measures to estimate this proportion as:

\begin{eqnarray}
Inf_{1}^{+}\left(u,v\right) & = & \textrm{Pr}_{u}^{\Theta}\left(Pos\right).m_{\left(u,v\right)}^{\Omega}\left(I\right)\label{eq:inf1}\\
Inf_{2}^{+}\left(u,v\right) & = & m_{u}^{\Theta}\left(Pos\right).m_{\left(u,v\right)}^{\Omega}\left(I\right)\label{eq:inf2}
\end{eqnarray}
In formula (\ref{eq:inf1}), we weight $m_{\left(u,v\right)}^{\Omega}\left(I\right)$
using $\textrm{Pr}_{u}^{\Theta}\left(Pos\right)$ to estimate the
proportion $Inf_{1}^{+}\left(u,v\right)$ \citep{Jendoubi2016a},
while in formula (\ref{eq:inf2}) we use $m_{u}^{\Theta}\left(Pos\right)$
to weight $m_{\left(u,v\right)}^{\Omega}\left(I\right)$ to estimate
$Inf_{2}^{+}\left(u,v\right)$. Indeed, formula (\ref{eq:inf2}) considers
the uncertainty of the user's opinion.

\subsubsection{Positive opinion influencers influencing positive users}

In the second scenario, we emphasize influence users who have a positive
opinion about the product and that are not connected to negative influencers.
In the first scenario, we defined two measures that estimate the positive
opinion influence of social users. In the second scenario, the goal
is to select among positive opinion influencers those that exert more
influence on positive users. For this purpose, we defined two measures
by weighting $Inf_{1}^{+}\left(u,v\right)$ and $Inf_{2}^{+}\left(u,v\right)$
using $\left(1-\textrm{Pr}_{v}^{\Theta}\left(Neg\right)\right)$ and
$\left(1-m_{v}^{\Theta}\left(Neg\right)\right)$ respectively as follows:

\begin{eqnarray}
Inf_{1}^{++}\left(u,v\right) & = & \textrm{Pr}_{u}^{\Theta}\left(Pos\right).m_{\left(u,v\right)}^{\Omega}\left(I\right).\left(1-\textrm{Pr}_{v}^{\Theta}\left(Neg\right)\right)\label{eq:inf++1}\\
 & = & Inf_{1}^{+}\left(u,v\right).\left(1-\textrm{Pr}_{v}^{\Theta}\left(Neg\right)\right)\\
Inf_{2}^{++}\left(u,v\right) & = & m_{u}^{\Theta}\left(Pos\right).m_{\left(u,v\right)}^{\Omega}\left(I\right).\left(1-m_{v}^{\Theta}\left(Neg\right)\right)\label{eq:inf++2}\\
 & = & Inf_{2}^{+}\left(u,v\right).\left(1-m_{v}^{\Theta}\left(Neg\right)\right)
\end{eqnarray}
The proposed measures, $Inf_{1}^{++}$ and $Inf_{2}^{++}$, give more
importance to the positive connection. Indeed, the values of $\left(1-\textrm{Pr}_{v}^{\Theta}\left(Neg\right)\right)$
and $\left(1-m_{v}^{\Theta}\left(Neg\right)\right)$ emphasize the
positive opinion of $u$'s neighbor. Consequently, multiplying the
positive influence that exerts $u$ on $v$ by the positive opinion
of $v$ ($\left(1-\textrm{Pr}_{v}^{\Theta}\left(Neg\right)\right)$
and $\left(1-m_{v}^{\Theta}\left(Neg\right)\right)$ respectively)
will result an influence measure that considers the positive opinion
of the influencer's neighbor.

\subsubsection{Positive opinion influencers influencing negative opinion users}

In the third scenario, we give more importance to influence users
who have a positive opinion about the product and that exert more
influence on negative influencers. Then, we define two influence measures
for this scenario by multiplying $Inf_{1}^{+}$ and $Inf_{2}^{+}$
with the non-positive proportion of $v$ opinion using respectively
$\left(1-\textrm{Pr}_{v}^{\Theta}\left(Pos\right)\right)$ and $\left(1-m_{v}^{\Theta}\left(Pos\right)\right)$
as:

\begin{eqnarray}
Inf_{1}^{+-}\left(u,v\right) & = & \textrm{Pr}_{u}^{\Theta}\left(Pos\right).m_{\left(u,v\right)}^{\Omega}\left(I\right).\left(1-\textrm{Pr}_{v}^{\Theta}\left(Pos\right)\right)\label{eq:inf+-1}\\
 & = & Inf_{1}^{+}\left(u,v\right).\left(1-\textrm{Pr}_{v}^{\Theta}\left(Pos\right)\right)\\
Inf_{2}^{+-}\left(u,v\right) & = & m_{u}^{\Theta}\left(Pos\right).m_{\left(u,v\right)}^{\Omega}\left(I\right).\left(1-m_{v}^{\Theta}\left(Pos\right)\right)\label{eq:inf+-2}\\
 & = & Inf_{2}^{+}\left(u,v\right).\left(1-m_{v}^{\Theta}\left(Pos\right)\right)
\end{eqnarray}
The proposed measures, $Inf_{1}^{+-}$ and $Inf_{2}^{+-}$, emphasize
negative connections. In fact, the values of $\left(1-\textrm{Pr}_{v}^{\Theta}\left(Pos\right)\right)$
and $\left(1-m_{v}^{\Theta}\left(Pos\right)\right)$ give more importance
to neighbors having a negative opinion about the product. Therefore,
multiplying the positive influence that exerts $u$ on $v$ by the
negative opinion of $v$ ($\left(1-\textrm{Pr}_{v}^{\Theta}\left(Pos\right)\right)$
and $\left(1-m_{v}^{\Theta}\left(Pos\right)\right)$ respectively)
will result an influence measure that considers the negative opinion
of the influencer's neighbor.

\subsection{Evidential influence maximization}

The influence spread function, $\sigma\left(S\right)$, is the function
to be maximized in the influence maximization process. It is the global
influence of a set of nodes, $S$, on all nodes in the social network.
To define $\sigma\left(S\right)$, first, we introduce the amount
of influence given to a set of nodes $S\subseteq V$ for influencing
a user, $v\in V$, in the network as follows:

\begin{equation}
\Phi\left(S,v\right)=\begin{cases}
1 & \textrm{if}\,v\in S\\
{\displaystyle \sum_{u\in S}{\displaystyle }\sum_{x\in D_{in}\left(v\right)\cup\left\{ v\right\} }Inf\left(u,x\right).Inf\left(x,v\right)} & \textrm{Otherwise}
\end{cases}\label{eq:infeq}
\end{equation}
such that $Inf\left(u,x\right)$ and $Inf\left(x,v\right)$ can be
estimated according to the marketing scenario prefixed by the marketer
using its corresponding measure as defined in (Section (\ref{subsec:User's-opinion-based})),
$Inf\left(v,v\right)=1$ and $D_{in}\left(v\right)$ is the set of
in-neighbors of $v$, \textit{i.e.} the set sources of directed links
having $v$ as a destination. 

Finally, we define the influence spread function $\sigma\left(S\right)$
under the evidential model as the total influence given to $S\subseteq V$
from all nodes in the social network as:

\begin{equation}
\sigma\left(S\right)=\sum_{v\in V}\Phi\left(S,v\right)
\end{equation}
In the spirit of the IM problem, as defined by \citep{Kempe03}, $\sigma\left(S\right)$
is the objective function to be maximized. Let us consider a social
network $G=\left(V,E\right)$, where $V$ is the set of nodes and
$E$ is the set of links and an objective function $\sigma\left(S\right)$
defined as explained above, then, find the set of nodes $S^{*}$ that
maximizes $\sigma\left(S\right)$ as follows:
\begin{equation}
S^{*}=\underset{S}{\textrm{argmax }}\sigma\left(S\right)
\end{equation}

To find out $S^{*}$, we used the influence maximization model that
was introduced in \citep{Jendoubi2017}. This model is the most adapted
for characteristic of the opinion based influence measures introduced
in this paper. The main goal of the influence maximization model is
to select a set of seed nodes $S^{*}$ that maximizes the objective
function $\sigma\left(S\right)$. Let $G=\left(V,\,E\right)$ be a
directed social network where $V$ is the set of nodes and $E$ is
the set of links and $k$ an integer where $k\leq|V|$. We note that
the maximization of $\sigma\left(S\right)$ was demonstrated to be
a NP-Hard problem \citep{Jendoubi2017}. Moreover, the influence spread
function was shown to be monotone and sub-modular in \citep{Jendoubi2017}.
All proof details can be found in \citep{Jendoubi2017}. To maximize
$\sigma\left(S\right)$, we use the cost effective lazy-forward algorithm
(CELF) \citep{Leskovec07b}. It is an extension of the greedy algorithm
that is proved to be about 700 times faster than the basic algorithm.
Next, we use the second influence maximization model that was introduced
in \citep{Jendoubi2017}. 

\section{Experiments\label{sec:Experiments}}

This section is mainly dedicated for experiments and results. In fact,
we evaluate the proposed influence measures on real world data and
generated data. Furthermore, we compare them to existing solutions
which are the credit distribution model and the opinion-based cascading
model (more details can be found in Section \ref{sec:Related-work}).
We consider the credit distribution model as base line model. This
solution was the first in the literature that considers the past propagation
to estimate the influence. The opinion-based cascading model is also
considered as a baseline model as it is the first model that considers
the user's opinion.

The real word data is used in this paper to evaluate the proposed
opinion-based influence measures. Then, we first study the quality
of the selected seeds using each measure. Besides, this dataset is
used to study the real opinion of the selected influencers. Next,
the generated data is used to study the ability of the influence maximization
model to detect the opinion based influencers for each scenario. 

\subsection{Data gathering and processing\label{sec:Data-gethering-and}}

In this section, we present the datasets we used in our experiments.
We also detail the process we followed and used tools to obtain our
data. Next, we propose two datasets, the first one was collected from
Twitter and the second one was randomly generated.

\subsubsection{Twitter dataset\label{subsec:Twitter-dataset}}

In our experiments, we define a Viral Marketing task which is about
the promotion of smartphones on Twitter. For this purpose, we crawled
Twitter data for the period between September, 8, 2014 and November,
3, 2014. We used the Twitter API through the Twitter4j java library\footnote{http://twitter4j.org/en/index.html}.
It is an open-sourced java implementation of the Twitter API, created
by Yusuke Yamamoto. Twitter API provides many kinds of data with some
limitations,\textit{ i.e}. a limited number of queries per hour or
limited response size. In our case, we are interested in collecting
tweets written in English, users, who mentions whom and who retweets
from whom. Next, we filtered the obtained data by keeping only tweets
that talk about smartphones and users having at least one tweet in
the data base. In a last step, we used the process explained in the
Section \ref{subsec:User's-opinion-estimation} to estimate the opinion
of each user in the dataset about smartphones. This dataset was also
used in the experiments of \citep{Jendoubi2016a,Jendoubi2017}. In
this dataset, we have no information about neither the influence users
nor the set of users that maximizes the influence. The influence maximization
using this dataset allows to evaluate the proposed opinion-based influence
measures in a real world case, \textit{i.e.} the influence users are
not known beforehand like in real world cases. Then, we use the proposed
solutions to estimate the influence, the passivity of each user and
to detect the influencers. Next, we evaluate the quality of the selected
influencers and their opinion which allows to evaluate the relevance
of the proposed scenarios and measures. 

\begin{sidewaysfigure}
\begin{centering}
\includegraphics[scale=0.4]{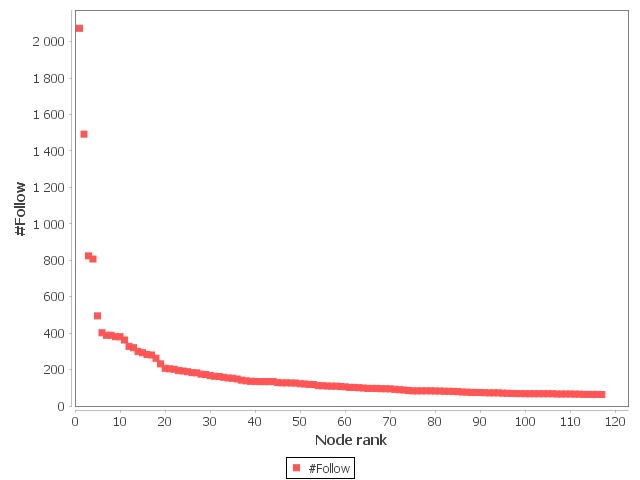} \includegraphics[scale=0.4]{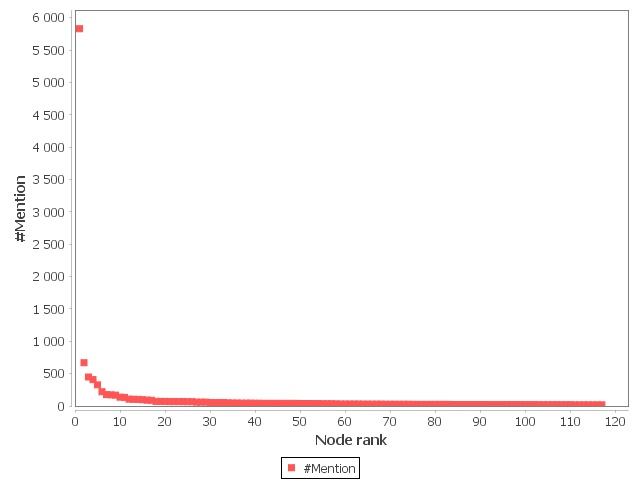}
\par\end{centering}
\begin{centering}
\includegraphics[scale=0.4]{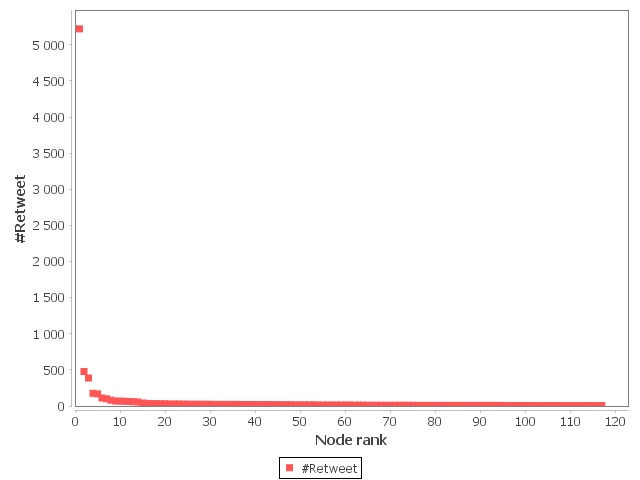} \includegraphics[scale=0.4]{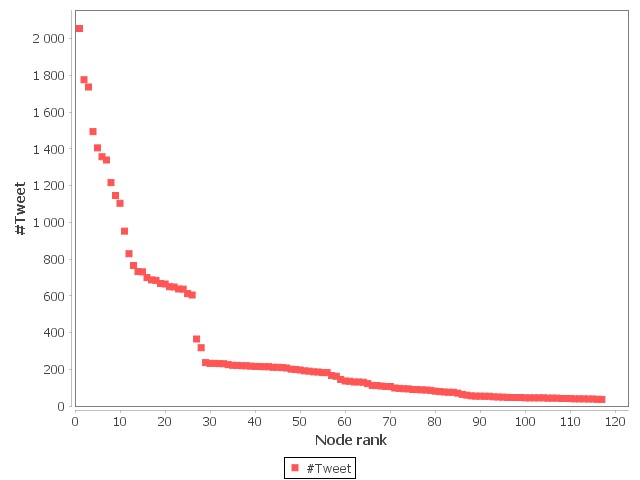}
\par\end{centering}
\caption{Data distributions \citep{Jendoubi2017} \label{fig:Data-distributions}}
\end{sidewaysfigure}

Table \ref{tab:Statistics-1} presents some statistics about the content
of the collected data. Besides, Figure \ref{fig:Data-distributions}
displays data distributions over users based on the number of followers,
mentions, retweets and tweets across our data. The follow relationship
is an explicit relation between Twitter user. In fact, when a user
$u$ follows another user $v$, $u$ will receive all the activity
of $v$. The mention and the retweet are implicit relations in Twitter.
Besides, these relations permit the information propagation on the
network. Finally, a tweet is 140 characters message. 

\begin{table}
\caption{Statistics of the dataset\label{tab:Statistics-1} \citep{Jendoubi2017}}
\centering{}%
\begin{tabular}{|c|c|c|c|c|}
\hline 
\textbf{\#User} & \textbf{\#Tweet} & \textbf{\#Follow} & \textbf{\#Retweet} & \textbf{\#Mention}\tabularnewline
\hline 
36,274 & 251,329 & 71,027 & 9,789 & 20,300\tabularnewline
\hline 
\end{tabular}
\end{table}

\subsubsection{Generated dataset\label{subsec:Generated-dataset}}

The generated data is used in this paper to study the performance
of the proposed influence measures. In fact, we generated data in
such a way one can know the influencers, the positive influencers,
the positive influencers influencing positive users and the positive
influencers influencing negative users. Indeed, to the best of our
knowledge, there is no available annotated dataset for the influence
maximization problem. Besides, annotating the real word dataset is
not given. Then comes the solution of the generated dataset to obtain
an annotated data that allows the evaluation of the proposed solution.
Next, we obtained a useful dataset to study the accuracy of the used
influence maximization solutions. Indeed, this dataset is useful to
study the adaptability of the used influence maximization model for
the problem of the opinion-based influence maximization and for the
defined measures. Next, we detail the process we used to obtain this
data. The proposed process is parameterizable and allows to study
the accuracy variation in terms of each parameter.

Social networks have special characteristics making them different
when compared to ordinary graphs. Among these characteristics, we
find the small world assumption \citep{Newman10}. For this reason,
we selected a random sampling from Twitter dataset (Section \ref{subsec:Twitter-dataset}).
The resulting network contains 1,010 vertices and 6,906 directed links.
Next, we defined a set of influence users. These users are chosen
according to their number of outlinks. Besides, this number is fixed
according to the assumption that the number of influencers is about
10\% of the total number of users in the network. Then, we defined
an influencer in this dataset to be a user having at least 15 outlinks.
As a result, 108 users are selected. In fact, taking a number of outlinks
less than 15 will lead to more selected users, and taking a number
greater than 15 will lead to a small number of influencers in this
experiment. Then, we consider this value the most adapted for the
network we have. 

In the next step, we defined randomly, the influence of each user
in the network by keeping the selected 108 users as top influencers.
Then, we assigned an influence value to each link in the network.
We fixed ``the minimum value of influence'' as a parameter of the
random process.  In a third step, we selected positive influencers
among the fixed set of influencers and we assigned to them a positive
opinion. The ``minimum value of positive opinion'' is a parameter
to the random process. In a last step, we defined among positive influencers
those that influences positive and negative users. For this purpose,
we divided the set of positive influencers into two random subsets.
The first subset is for positive influencers influencing positive
users, then, we set the opinion of the influencers neighbors to positive.
The second subset is for positive influencers influencing negative
users and we set the opinion of their neighbors to negative. We notice
that there are two more parameters of the random process which are
the ``minimum positive and negative opinion of positive influencers'
neighbors''.

Using the defined random process, we assign to each link in the network
three random values which are the influence, the positive opinion
and the negative opinion. The random opinions are used to define influencers
for each scenario. We note that we consider a user more influencer
when his influence value is near 1. Then, we fix the \textquotedblleft minimum
value of influence\textquotedblright{} to study the behavior of the
maximization model when the influence increases which gives an idea
about realistic cases. However, fixing the maximum value of influence
is not realistic. Indeed, when the maximum is near 0, then there is
no influencer in the network. Besides, when the maximum is near to
1, then the system will consider all the users in the network as influencers.

\subsection{Impact of the opinion incorporation}

In this section, we perform some experiments to study the impact of
the user's opinion about the product on the detected seeds. The task
we propose is about the influence maximization to promote smartphones
on Twitter. The main purpose of this task is to find a set of $k$
influencer users that are able to maximize the global positive influence
through the network and to promote the adoption of smartphones. We
use the dataset collected from Twitter, the reader can refer to Section
\ref{subsec:Twitter-dataset} for more details. Let us define $Inf\left(u,v\right)=m_{\left(u,v\right)}^{\Omega}\left(I\right)$
to be the evidential influence measure that does not consider the
user's opinion. An estimation process of this measure was introduced
in \citep{Jendoubi2017}. Furthermore, we use the second evidential
model introduced in \citep{Jendoubi2017} with the proposed influence
measures as follows:
\begin{itemize}
\item The evidential influence measure, $Inf$, ($Inf\left(u,v\right)=m_{\left(u,v\right)}^{\Omega}\left(I\right)$),
called ``Evidential model''. This model and evidential measure were
presented in detail in \citep{Jendoubi2017}.
\item The first measure of the first scenario, $Inf_{1}^{+}$, (equation
(\ref{eq:inf1})), called ``First scenario with probability opinion''.
\item The second measure of the first scenario, $Inf_{2}^{+}$, (equation
(\ref{eq:inf2})), called ``First scenario with belief opinion''.
\item The first measure of the second scenario, $Inf_{1}^{++}$, (equation
(\ref{eq:inf++1})), called ``Second scenario with probability opinion''.
\item The second measure of the second scenario, $Inf_{2}^{++}$, (equation
(\ref{eq:inf++2})), called ``Second scenario with belief opinion''.
\item The first measure of the third scenario, $Inf_{1}^{+-}$, (equation
(\ref{eq:inf+-1})), called ``Third scenario with probability opinion''.
\item The second measure of the third scenario, $Inf_{2}^{+-}$, (equation
(\ref{eq:inf+-2})), called ``Third scenario with belief opinion''.
\end{itemize}
In the following experiments, we fixed $k$ to $50$ and we compared
the proposed solutions to the Credit Distribution model (CD) \citep{Goyal12}
and to the Opinion-based Cascading (OC) model \citep{Zhang2013}.
We note that we adapted the CD model to Twitter data by defining two
actions which are the mention and the retweet. Then, we estimated
the credit of each user in the network using these two actions as
defined in the algorithm (see Section \ref{subsec:Influence-maximization}).
Furthermore, we used the estimated opinion values (used in the proposed
measures) to define the function $Op\left(v\right)$ of the OC model
(see Section \ref{subsec:Opinion-based-influence-maximiza} for more
details about OC). Each of these two models has some common properties
with the proposed influence maximization solutions. In fact, the credit
distribution influence measure uses past propagation to estimate the
influence as the proposed measures, but it does not consider the opinion.
Besides, OC considers the user's opinion in its influence maximization
process. However, it does not use past propagation to estimate the
influence. 

In a first experiment, we compare the number of common selected seeds
on Twitter dataset as shown in Table \ref{tab:Seed-sets-intersection}.
Then, we examine the amount of common selected seeds between each
two couples of models. Indeed, the number of common selected seeds
can be seen as a similarity indicator between influence maximization
models. Then, it shows if there are some similarities between the
experimented models. Furthermore, this experiment is important as
it allows to understand the extent to which the user's opinion has
an impact on the set of selected seeds. We notice that the Opinion-based
Cascading model (OC) has no common seeds with any experienced model.
Besides, CD model has no more than nine common seeds with ``Evidential
model'' that uses an influence measure on the set $\left\{ Inf_{1}^{+},\,Inf_{2}^{+},\,Inf_{1}^{++},\,Inf_{2}^{++}\right\} $.
However, CD has only one common seed with ``Evidential model'' with
$\left\{ Inf_{1}^{+-},\,Inf_{2}^{+-},\,Inf\right\} $. Furthermore,
``Evidential model'' has a little number of common seeds with other
experienced models. However, we have at least 34 common seeds between
any couple of models using any influence measure from the set $\left\{ Inf_{1}^{+},\,Inf_{2}^{+},\,Inf_{1}^{++},\,Inf_{2}^{++}\right\} $.
Besides, the ``third scenario with belief opinion'' and ``third
scenario with probability opinion'' have 47 common seeds. We explain
these observations by the fact that the used opinion-based influence
measures are similar because all of them are based on the evidential
influence measure, $Inf$. 

\begin{sidewaystable}
\caption{Seed sets intersection\label{tab:Seed-sets-intersection}}
\centering{}%
\begin{tabular}{|>{\centering}m{20mm}|>{\centering}m{14mm}|>{\centering}m{14mm}|>{\centering}m{16mm}|>{\centering}m{14mm}|>{\centering}m{17mm}|>{\centering}m{16mm}|>{\centering}m{17mm}>{\centering}m{16mm}>{\centering}m{17mm}}
\cline{2-10} 
\multicolumn{1}{>{\centering}m{20mm}|}{} & {\small{}OC} & {\small{}CD} & {\small{}Evidential model} & {\small{}Third scenario with belief opinion} & {\small{}Third scenario with probability opinion} & {\small{}Second scenario with belief opinion} & \multicolumn{1}{>{\centering}m{17mm}|}{{\small{}Second scenario with probability opinion}} & \multicolumn{1}{>{\centering}m{16mm}|}{{\small{}First scenario with belief opinion}} & \multicolumn{1}{>{\centering}m{17mm}|}{{\small{}First scenario with probability opinion}}\tabularnewline
\hline 
{\small{}First scenario with probability opinion} & {\small{}0} & {\small{}9} & {\small{}9} & {\small{}17} & {\small{}17} & {\small{}40} & \multicolumn{1}{>{\centering}m{17mm}|}{{\small{}35}} & \multicolumn{1}{>{\centering}m{16mm}|}{{\small{}38}} & \multicolumn{1}{>{\centering}m{17mm}|}{{\small{}50}}\tabularnewline
\hline 
{\small{}First scenario with belief opinion} & {\small{}0} & {\small{}7} & {\small{}7} & {\small{}15} & {\small{}15} & {\small{}34} & \multicolumn{1}{>{\centering}m{17mm}|}{{\small{}42}} & \multicolumn{1}{>{\centering}m{16mm}|}{{\small{}50}} & \tabularnewline
\cline{1-9} 
{\small{}Second scenario with probability opinion} & {\small{}0} & {\small{}9} & {\small{}9} & {\small{}18} & {\small{}18} & {\small{}40} & \multicolumn{1}{>{\centering}m{17mm}|}{{\small{}50}} &  & \tabularnewline
\cline{1-8} 
{\small{}Second scenario with belief opinion} & {\small{}0} & {\small{}8} & {\small{}8} & {\small{}18} & {\small{}18} & {\small{}50} &  &  & \tabularnewline
\cline{1-7} 
{\small{}Third scenario with probability opinion} & {\small{}0} & {\small{}1} & {\small{}16} & {\small{}47} & {\small{}50} & \multicolumn{1}{>{\centering}m{16mm}}{} &  &  & \tabularnewline
\cline{1-6} 
{\small{}Third scenario with belief opinion} & {\small{}0} & {\small{}1} & {\small{}13} & {\small{}50} & \multicolumn{1}{>{\centering}m{17mm}}{} & \multicolumn{1}{>{\centering}m{16mm}}{} &  &  & \tabularnewline
\cline{1-5} 
{\small{}Evidential model} & {\small{}0} & {\small{}1} & {\small{}50} & \multicolumn{1}{>{\centering}m{14mm}}{} & \multicolumn{1}{>{\centering}m{17mm}}{} & \multicolumn{1}{>{\centering}m{16mm}}{} &  &  & \tabularnewline
\cline{1-4} 
{\small{}CD}\\
 & {\small{}0} & {\small{}50} & \multicolumn{1}{>{\centering}m{16mm}}{} & \multicolumn{1}{>{\centering}m{14mm}}{} & \multicolumn{1}{>{\centering}m{17mm}}{} & \multicolumn{1}{>{\centering}m{16mm}}{} &  &  & \tabularnewline
\cline{1-3} 
{\small{}OC}\\
 & {\small{}50} & \multicolumn{1}{>{\centering}m{14mm}}{} & \multicolumn{1}{>{\centering}m{16mm}}{} & \multicolumn{1}{>{\centering}m{14mm}}{} & \multicolumn{1}{>{\centering}m{17mm}}{} & \multicolumn{1}{>{\centering}m{16mm}}{} &  &  & \tabularnewline
\cline{1-2} 
\end{tabular}
\end{sidewaystable}

In a second experiment, we compare the mean positive and negative
opinions of the selected seeds and their neighbors using each maximization
model on Twitter dataset as shown in Table \ref{tab:Seeds-mean-opinion}.
This experiment allows to evaluate each experimented model in terms
of the opinion of selected seeds and their neighbors. Indeed, we seek
influencers according to their positive opinions and to the opinion
of their neighbors, then it is interesting to verify if this condition
was satisfied in the selected seed set. In Table \ref{tab:Seeds-mean-opinion},
``Evidential model'' that uses the evidential influence measure,
$Inf$, selects influencer spreaders that have a moderate positive
and negative opinion, about $0.3$. This fact is expected, because
$Inf$ does not consider the user's opinion. Besides, the CD model
chooses influencers that have a small value of positive and negative
opinion, about $0.01$, which proves that it is not adaptable for
this purpose. In fact, CD does not consider the user's opinion \citep{Jendoubi2016a}.
However, the OC model selects seeds with about $0.41$ of mean positive
opinion which still an unsatisfactory result for a model that considers
the user's opinion about the product. 

In another hand, when we consider the user's opinion, we notice better
results in the ``mean positive opinion'' and ``mean negative opinion''
of selected seeds. Indeed, models that use an influence measure from
the set\\
 $\left\{ Inf_{1}^{+},\,Inf_{2}^{+},\,Inf_{1}^{++},\,Inf_{2}^{++}\right\} $,
select seeds having at least about $0.82$ of ``mean positive opinion''
and at most about $0.08$ ``mean negative opinion'' which is an
impressive result compared to the results of existing models (CD and
OC). In Table \ref{tab:Seeds-mean-opinion}, we notice that the best
maximization model in terms of mean positive and negative opinion
is the ``First scenario with belief opinion''. In fact, it gives
a maximum value of ``mean positive opinion'', which equals to $0.85\pm0.06$
($0.95$ confidence interval), and a minimum value of ``mean negative
opinion'', which equals to $0.06\pm0.03$. Furthermore, we observe
that all results of ``Evidential model'' with any influence measure
from the set $\left\{ Inf_{1}^{+},\,Inf_{2}^{+},\,Inf_{1}^{++},\,Inf_{2}^{++}\right\} $
are very near to each other's, this observation is explained in Table
\ref{tab:Seed-sets-intersection} where we find that they have many
common seeds.

The neighbors positive and negative opinion are now considered. We
find that the best ``mean positive opinion'' value of seeds neighbors
is given by ``Evidential model'' with $Inf_{1}^{++}$ and $Inf_{2}^{++}$.
Indeed, we have got $0.42\pm0.07$ which is the highest value compared
to those given by the other proposed influence measures, CD and OC
models. This observation can be explained by the fact that $Inf_{1}^{++}$
and $Inf_{2}^{++}$ consider the positive opinion of the user's neighbors
while estimating the influence. In the same way, we notice that $Inf_{1}^{+-}$
and $Inf_{2}^{+-}$ detect seeds with highest ``mean negative opinion''
of seed's neighbors. In fact, they give a value of $0.22\pm0.06$
which is the maximum value in the last column of the Table \ref{tab:Seeds-mean-opinion}.
This fact is explained by the consideration of the negative opinion
of the user's neighbors while estimating the influence.

\begin{table}
\caption{Mean opinions of selected seeds and their neighbors\label{tab:Seeds-mean-opinion}}
\centering{}%
\begin{tabular}{|>{\centering}p{2cm}|>{\centering}p{25mm}|>{\centering}p{25mm}|>{\centering}p{25mm}|>{\centering}p{25mm}|}
\hline 
{\small{}Model} & {\small{}Mean positive opinion} & {\small{}Mean negative opinion} & {\small{}Mean positive neighbors opinion} & {\small{}Mean negative neighbors opinion}\tabularnewline
\hline 
{\small{}First scenario with probability opinion} & {\small{}$0.83\pm0.07$} & {\small{}$0.07\pm0.03$} & {\small{}$0.39\pm0.06$} & {\small{}$0.21\pm0.06$}\tabularnewline
\hline 
{\small{}First scenario with belief opinion} & {\small{}$0.85\pm0.06$} & {\small{}$0.06\pm0.03$} & {\small{}$0.40\pm0.06$} & {\small{}$0.21\pm0.06$}\tabularnewline
\hline 
{\small{}Second scenario with probability opinion} & {\small{}$0.80\pm0.07$} & {\small{}$0.08\pm0.03$} & {\small{}$0.42\pm0.07$} & {\small{}$0.20\pm0.07$}\tabularnewline
\hline 
{\small{}Second scenario with belief opinion} & {\small{}$0.82\pm0.07$} & {\small{}$0.07\pm0.03$} & {\small{}$0.42\pm0.07$} & {\small{}$0.20\pm0.07$}\tabularnewline
\hline 
{\small{}Third scenario with probability opinion} & {\small{}$0.59\pm0.06$} & {\small{}$0.18\pm0.03$} & {\small{}$0.39\pm0.06$} & {\small{}$0.22\pm0.06$}\tabularnewline
\hline 
{\small{}Third scenario with belief opinion} & {\small{}$0.62\pm0.06$} & {\small{}$0.17\pm0.03$} & {\small{}$0.39\pm0.06$} & {\small{}$0.22\pm0.06$}\tabularnewline
\hline 
{\small{}Evidential model} & {\small{}$0.30\pm0.04$} & {\small{}$0.30\pm0.02$} & {\small{}$0.39\pm0.06$} & {\small{}$0.21\pm0.06$}\tabularnewline
\hline 
{\small{}CD}\\
 & {\small{}$0.01\pm0.02$} & {\small{}$0.01\pm0.01$} & {\small{}$0.24\pm0.05$} & {\small{}$0.13\pm0.05$}\tabularnewline
\hline 
{\small{}OC}\\
 & {\small{}$0.41\pm0.08$} & {\small{}$0.20\pm0.04$} & {\small{}$0.35\pm0.08$} & {\small{}$0.21\pm0$}\tabularnewline
\hline 
\end{tabular}
\end{table}

In a last experiment, the purpose is to examine the quality of the
selected influencers for smartphones on Twitter. Then, we fix a set
of comparison criteria. Indeed, we choose the accumulated number of
followers, \textit{\#Follow}, the accumulated number of times the
user was mentioned, \textit{\#Mention}, the accumulated number of
times the user was retweeted,\textit{ \#Retweet} and the accumulated
number of tweets, \textit{\#Tweet}. In fact, if a given user is an
influencer on Twitter, he is necessarily: very active then he has
a lot of tweets, he is followed by many users in the network that
are interested in his news, he is frequently mentioned in other tweets
and his tweets are retweeted several times. These assumptions justify
the chosen comparison criteria. 

\begin{sidewaysfigure}
\begin{centering}
\includegraphics[scale=0.45]{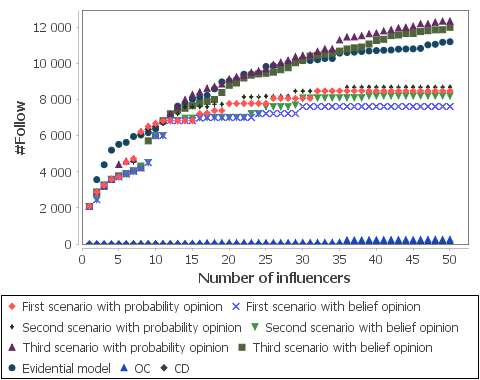}
\includegraphics[scale=0.45]{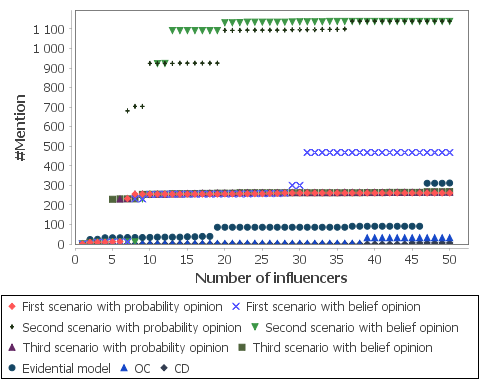}
\par\end{centering}
\begin{centering}
\includegraphics[scale=0.45]{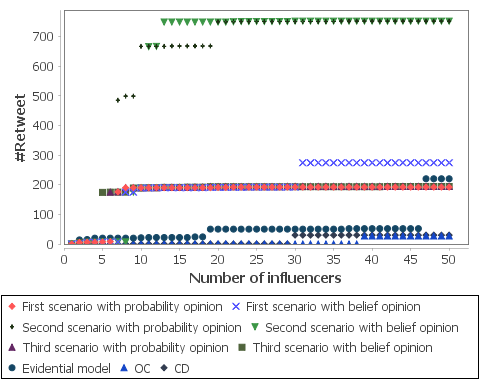}
\includegraphics[scale=0.45]{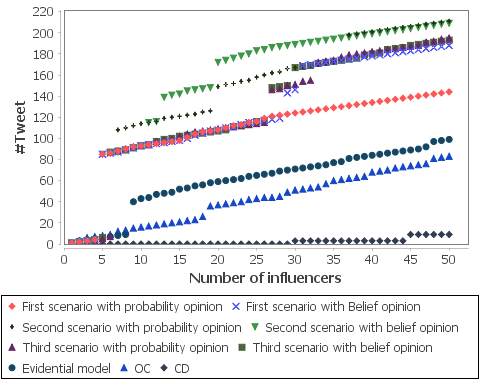}
\par\end{centering}
\caption{Comparison between the opinion based scenarios, the second influence
model and the OC model\label{fig:Comparision-between}}
\end{sidewaysfigure}

We compare all experimented models in terms of \#Follow, \#Mention,
\#Retweet and \#Tweet. This experiment is useful to study and compare
the quality of selected seeds on Twitter dataset using each experimented
model. As a result, we have got curves presented in Figure \ref{fig:Comparision-between}. 

In Figure \ref{fig:Comparision-between}, we have four sub-figures
in which we present the accumulated \#Follow, \#Mention, \#Retweet
and \#Tweet respectively. In the accumulated \#Follow figure, we notice
that all experimented models selected seeds that are followed by many
other users except CD and OC models that their seeds do not exceed
ten followers in all. Besides, the ``Evidential model'', the ``Third
scenario with probability opinion'' and the ``Third scenario with
belief opinion'' give almost the same results that are up to 12,000
accumulated \#Follow. Furthermore, the results of the first and the
second scenarios are very similar and are up to about 9,000 accumulated
\#Follow.

In a second sub-Figure of Figure \ref{fig:Comparision-between}, we
have the accumulated \#Mention curves. We observe that OC and CD models
do not select mentioned seeds and their accumulated \#Mention values
do not exceed twenty in all. Furthermore, the ``Evidential model''
that uses an opinion-based measure (the three scenarios) has better
results in terms of accumulated \#Mention than ``Evidential model''
that uses the evidential influence measure $Inf$. Besides, the ``second
scenario with probability opinion'' and the ``second scenario with
belief opinion'' have the best results between all the experimented
models in terms of accumulated \#Mention. Indeed, they reach over
1100 \#Mention from about the twentieth selected seed. From the results
of this sub-Figure, we can conclude that the incorporation of the
user's opinion in the process of the influence maximization ameliorates
the quality of selected seeds in terms of accumulated \#Mention.

In a third sub-Figure of Figure \ref{fig:Comparision-between}, we
study the quality of selected seeds by each experimented model in
terms of accumulated \#Retweet. We observe that selected seeds using
OC or CD models are not retweeted a lot. In fact, their curves do
not exceed fifty accumulated \#Retweet. In addition, we notice a similar
behavior of the proposed influence measures to the accumulated \#Mention
curves. In fact, we see that the three proposed scenarios have succeeded
in selecting seeds having a high accumulated \#Retweet. Also, the
``second scenario with probability opinion'' and the ''second scenario
with belief opinion'' give the best results in terms of accumulated
\#Retweets. 

In the last sub-Figure, we study the quality of the selected seeds
in terms of accumulated \#Tweet on Twitter dataset. In this sub-Figure
we observe a different behavior of OC model. In fact, it succeeds
in selecting some active users in terms of accumulated \#Tweet. However,
it does not reach the activity level of seeds detected by the proposed
influence maximization solutions. Besides, we notice that CD model
does not exceed twenty accumulated \#Tweet in all. In another hand,
we notice that the proposed influence maximization solutions have
the same shape. Also, the ``second scenario with probability opinion''
and the ``second scenario with belief opinion'' detect the best
seeds in terms of accumulated \#Tweet. Besides, we observe that curves
of ``Evidential model'' with an opinion-based influence measure
exceeds the curve of ``Evidential model'' without considering the
opinion.

In Table \ref{tab:Running-time}, we have the running time in seconds
of the ``Evidential model'', ``First scenario with probabilistic
opinion'', CD and OC. This running time corresponds to the previous
experiments on the Twitter dataset. We notice that OC has the best
running time. Furthermore, the ``Evidential model'' and ``First
scenario with probabilistic opinion'' have almost the same running
time which is less than 5 seconds. Finally, CD model is the laziest
algorithm as it gives its results after 269.8 seconds.

\begin{table}
\caption{Running time in seconds\label{tab:Running-time}}
\centering{}%
\begin{tabular}{|c|>{\centering}p{2cm}|>{\centering}p{3cm}|c|c|}
\hline 
Model & Evidential model & First scenario with probabilistic opinion & CD & OC\tabularnewline
\hline 
Time & 4.7 & 4.3 & 269.8 & 1\tabularnewline
\hline 
\end{tabular}
\end{table}

In this section, we presented some interesting experiments using the
Twitter dataset to study and evaluate the proposed influence maximization
solutions. Furthermore, we defined the task of detecting influencers
for smartphones on Twitter. Our experiments show the performance of
the proposed opinion based influence measures in detecting good seeds
for smartphones. In fact, we notice a good improvement in the quality
of selected seeds not only in terms of the opinion about the product
but also in terms of \#Follow, \#Mention, \#Retweet and \#Tweet compared
to the ``Evidential model''. In the next section, we present a set
of experiments to study the accuracy of the proposed approach. 

\subsection{Adaptability of the used influence maximization model\label{sec:Studing-the-influence}}

In this section, we use the generated dataset introduced in Section
\ref{subsec:Generated-dataset} in order to study the behavior of
the proposed influence maximization solution while varying influence
and user opinion. The main purpose of these experiments is to study
the adaptability of the used influence maximization model and to justify
its choice. In these experiments, we fix the size of the seed set
$k$ to 50 and we repeat the random process twenty times. As we said
above in Section \ref{subsec:Generated-dataset}, the process used
to generate the data is parameterizable. Then, in our experiments,
we vary each parameter and we fix the others to study the accuracy
of the proposed influence maximization solutions. Finally, we experience
the following influence measures with the ``Evidential model'': 
\begin{itemize}
\item The evidential influence measure, $Inf$, ($Inf\left(u,v\right)=m_{\left(u,v\right)}^{\Omega}\left(I\right)$),
called ``Evidential model'', 
\item The first measure of the first scenario, $Inf_{1}^{+}$, (equation
(\ref{eq:inf1})), called ``First scenario with probability opinion'', 
\item The first measure of the second scenario, $Inf_{1}^{++}$, (equation
(\ref{eq:inf++1})), called ``Second scenario with probability opinion'', 
\item The first measure of the third scenario, $Inf_{1}^{+-}$, (equation
(\ref{eq:inf+-1})), called ``Third scenario with probability opinion''.
\end{itemize}
The main goal behind this section is to study the adaptability of
the used influence maximization model through a standard metric which
is the accuracy, defined as follows: 
\begin{equation}
Accuracy=\frac{NbrGoodPredicted}{NbrClassified}
\end{equation}
where $NbrGoodPredicted$ is the number of detected good influencers
according to the used influence measure, $Inf$, $Inf_{1}^{+}$, $Inf_{1}^{++}$
and $Inf_{1}^{+-}$ respectively. $NbrClassified$ is the total number
of the influencers defined for each category respectively. Besides,
in these experiments we did not consider the CD and OC models. Indeed,
these two models do not use a similar influence estimation process
as the proposed solutions. Then, the random generated influence values
can not be used with these two models.

\begin{figure}
\begin{centering}
\subfloat[Accuracy of detected influencers while varying the minimum influence
value\label{fig:Accuracy-Inf_VarInf}]{\begin{centering}
\includegraphics[scale=0.55]{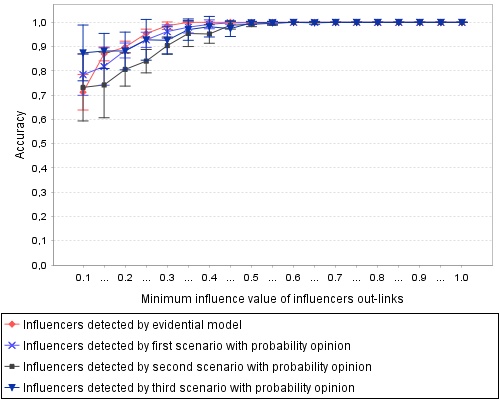}
\par\end{centering}
}
\par\end{centering}
\begin{centering}
\subfloat[Accuracy of detected positive influencers while varying the minimum
influence value\label{fig:Accuracy-Pos_VarInf}]{\begin{centering}
\includegraphics[scale=0.55]{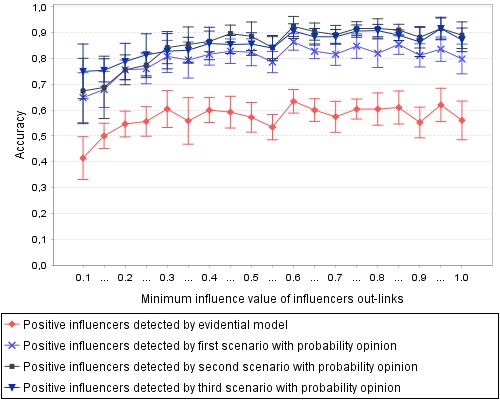}
\par\end{centering}
}
\par\end{centering}
\caption{Accuracy variation while varying the minimum influence value\label{fig:Accuracy-variation}}
\end{figure}

In a first experiment, we use the generated dataset and we vary the
minimum influence parameter and we study its impact on detecting influencers
and positive influencers as shown in Figure \ref{fig:Accuracy-variation}.
We fix the minimum positive opinion of positive influencers to 0.8,
the minimum positive and negative opinion of positive influencers
neighbors to 0.3 and 0.8 respectively. Figure \ref{fig:Accuracy-Inf_VarInf}
shows the accuracy of detecting influencers by the experimented models
while varying the minimum influence value. This figure shows the performance
of the proposed models. In fact, even with a small influence value,
0.1, the experimented models succeed in detecting influencers with
a good accuracy that is no less than 80\%. In fact, when the influence
value is small, there is more confusion between network users and
influencers. In this case, we notice that our system is able to manage
this confusion. Besides, we notice that the ``Evidential model''
starts having the highest accuracy from the influence value 0.15 until
the value 0.6 from where all other models start having accuracy equal
to 1. 

In Figure \ref{fig:Accuracy-Pos_VarInf}, we study the accuracy of
detecting positive influencers while varying the ``minimum influence''
value. In this figure, we observe that the first, the second and the
third scenarios give good accuracy of detecting seeds having a positive
opinion. Besides, we notice a natural behavior of the ``Evidential
model'' that does not consider the opinion in its principle, but
it keeps giving acceptable accuracy.

\begin{figure}
\begin{centering}
\includegraphics[scale=0.55]{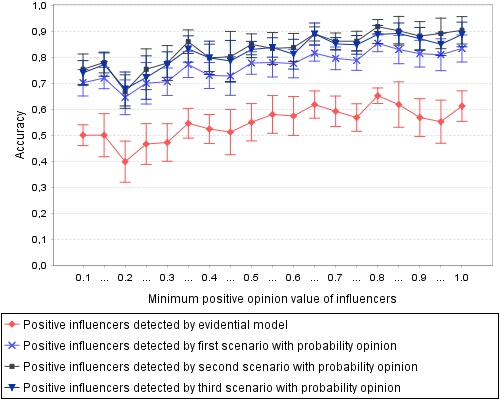}
\par\end{centering}
\caption{Accuracy of detected positive influencers while varying the minimum
positive opinion value\label{fig:Accuracy-Pos_VarPos}}
\end{figure}

In a second experiment, we vary the ``minimum positive opinion''
value of influence users in the generated data. In this experiment,
we fixed the minimum influence value to 0.5, the minimum positive
and negative opinion of positive influencers neighbors to 0.3 and
0.8 respectively. Figure \ref{fig:Accuracy-Pos_VarPos} presents the
accuracy of detecting influencers having a positive opinion by the
mean of each experimented model. In this figure, we notice a similar
results to those of the Figure \ref{fig:Accuracy-Pos_VarInf}. In
fact, all curves are almost steady. Besides, the best accuracy is
given by the second, the third and the first scenarios respectively. 

\begin{figure}
\begin{centering}
\includegraphics[scale=0.55]{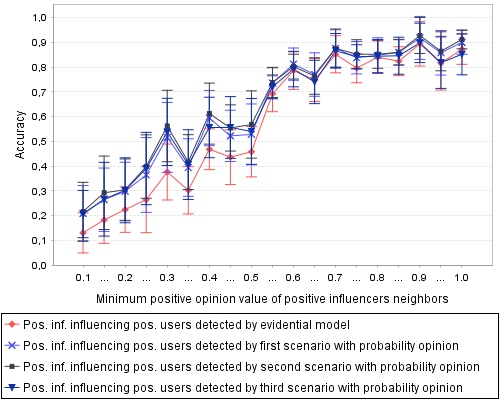}
\par\end{centering}
\caption{Accuracy of detected positive influencers influencing positive users
while varying the minimum positive opinion value of positive influencers
neighbors\label{fig:Accuracy-PosPos_VarPosPos}}
\end{figure}

In a third experiment, we vary the ``minimum positive opinion of
positive influencers neighbors'' in the generated data. In this experiment,
we fixed the minimum influence value to 0.4, the minimum positive
opinion of influencers to 0.5 and the minimum negative opinion of
positive influencers neighbors to 0.8. Figure \ref{fig:Accuracy-PosPos_VarPosPos}
shows the accuracy of detecting positive influencers that exert more
influence on positive users. In this figure, we notice a different
behavior from the previous figures. In fact, all curves increase gradually
when the minimum positive opinion value of positive influencers neighbors
increases until getting high accuracy.

\begin{figure}
\begin{centering}
\subfloat[Accuracy of detected positive influencers influencing negative users
while varying the minimum negative opinion value of positive influencers
neighbors\label{fig:Accuracy-PosNeg_VariationPosNeg}]{\begin{centering}
\includegraphics[scale=0.55]{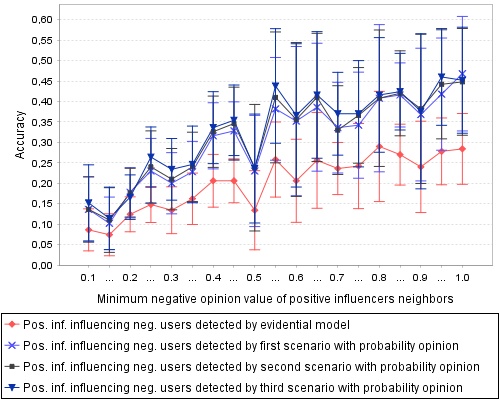}
\par\end{centering}
}
\par\end{centering}
\begin{centering}
\subfloat[Accuracy of detected positive influencers influencing positive users
while varying the minimum negative opinion value of positive influencers
neighbors\label{fig:Accuracy-PosPos_VariationPosNeg}]{\begin{centering}
\includegraphics[scale=0.55]{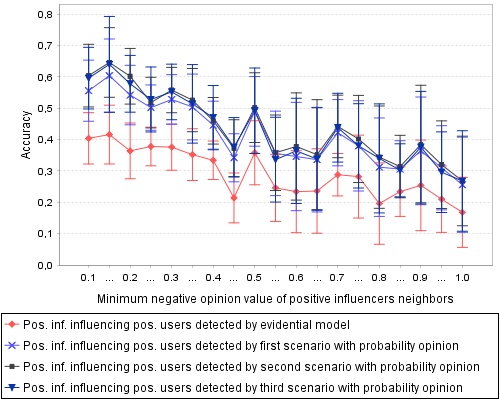}
\par\end{centering}
}
\par\end{centering}
\caption{Accuracy of detected positive influencers influencing positive and
negative users while varying the minimum negative opinion value of
positive influencers neighbors\label{fig:Accuracy-VariationPosNeg}}
\end{figure}

In a last experiment, we vary the ``minimum negative opinion of positive
influencers neighbors'' in the generated data. Besides, we fix the
minimum influence value to 0.4, the minimum positive opinion of influencers
to 0.5 and the minimum positive opinion of positive influencers neighbors
to 0.2. Figure \ref{fig:Accuracy-VariationPosNeg} shows the accuracy
of detecting positive influencers that exert more influence on positive
and negative users while varying the minimum negative opinion value
of positive influencers neighbors. In the first sub-figure \ref{fig:Accuracy-PosNeg_VariationPosNeg},
we have the accuracy of detected positive influencers influencing
negative users. All curves increase when the varied value increases.
Besides, the best accuracy values are given by the third scenario
which is dedicated to positive influencers that exert more influence
on negative users. In the second sub-figure \ref{fig:Accuracy-PosPos_VariationPosNeg},
we have the accuracy of detected positive influencers influencing
positive users. In this figure, we observe a reverse behavior of curves
in Figure \ref{fig:Accuracy-PosNeg_VariationPosNeg}. In fact, the
accuracy decreases when the varied value increases. This behavior
is explained by the fact that, when the number of positive influencers
influencing negative users increases, the number of those influencing
positive users decreases.

To sum up, in this section, we presented some results made on generated
data. These results show the adaptability of the used influence maximization
model to detect opinion based influencers. Besides, we notice that
the results of the first, the second and the third scenarios are similar.
This behavior is justified by the similarity between their influence
measures. 

\section{Conclusion}

In this paper, we mainly focus on the importance of the user's opinion
when it comes to measuring and maximizing the social influence. In
fact, the opinion is a crucial parameter that can determine whether
the success or the failure of a viral marketing campaign. To make
such a campaign more successful, we introduce three opinion-based
viral marketing scenarios. The first scenario is called ``positive
influencers'', its main aim is to find social influencers having
a positive opinion about the product. The second scenario is ``positive
opinion influencers influencing positive users'', this scenario is
about positive influencers that exert more influence on users having
a positive opinion. The third scenario is ``positive opinion influencers
influencing negative opinion users'', it looks for positive influencers
that exert more influence on users having a negative opinion. For
each scenario, we introduced two appropriate influence measures. Furthermore,
we used these measures to maximize the influence. For this purpose,
we used an adapted maximization model which is the evidential influence
model \citep{Jendoubi2017}. Next, we run a set of experiments to
show the performance of the proposed solutions. Then, we studied the
quality of the detected influencers and their positive and negative
opinions using a real world dataset from Twitter network. Besides,
we studied the adaptability of the used maximization model to detect
the adapted influencers for each scenario and we showed that through
a set of experiments on a generated dataset.

An interesting perspective for future works is about maximizing the
influence within communities. A community is defined as a set of users
or vertices that are connected more densely to each other than to
other users from other communities \citep{Narayanam2014,Zhouab2015,Moosavi2017}.
People in the same community generally have some common properties.
For example, they may be friends that attended the same school or
they are from the same town. The idea here is to minimize the number
of selected influencers and the time spent to find them. In fact,
we will search to find influencers at the scale of the community instead
of the social network and it is obvious that the community is smaller
than the social network.

\bibliographystyle{splncs03}
\bibliography{biblio}
\newpage

\begin{figure}[h]
	\centering
		\includegraphics[width=.20\textwidth]{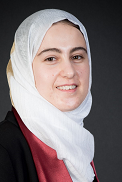}
	\label{fig:Author1}
\end{figure}
Siwar Jendoubi is a researcher at LARODEC Laboratory. She received a PhD degree (2016) in computer science and a Master degree (2012) in business intelligence. Dr. Siwar Jendoubi joined the LISTIC Laboratory for 18 months as postdoctoral researcher. Her research interests are mainly related to information fusion, uncertainty management, machine learning, social network analysis and tree species recognition. She is author of many papers.

\begin{figure}[h]
	\centering
		\includegraphics[width=.20\textwidth]{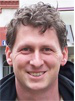}
	\label{fig:Author2}
\end{figure}
Arnaud Martin is full professor at University of Rennes 1 in the team DRUID of IRISA laboratory. He received a HDR (French ability to supervised research) in computer sciences (2009), a PhD degree in Signal Processing (2001), and Master in Probability (1998). Pr. Arnaud Martin joined the laboratory IRISA at the university of Rennes 1 as full professor in 2010 and co-create the team DRUID in 2012. He teaches data fusion, data mining, and computer sciences. His research interests are mainly related to the belief functions with applications on social networks and crowdsourcing. He is author of numerous papers and invited talks. He supervised numerous Phd students. Pr. Arnaud Martin is the founder in 2010 of the Belief Functions and Applications Society (BFAS) (www.bfasociety.org). From 2010 until 2012 he was the president of BFAS, and since 2012 he is the treasurer. 

\end{document}